\documentclass[twocolumn,english,aps,manuscript,prl,reprint,longbibliography]{revtex4-1}
\usepackage[T1]{fontenc}
\usepackage[latin9]{inputenc}
\usepackage{amsmath}
\usepackage{amssymb}
\usepackage{graphicx}
\usepackage{natbib}

\makeatletter

\usepackage[usenames,dvipsnames]{color}
\@ifundefined{textcolor}{}
{%
\definecolor{BLACK}{gray}{0}
\definecolor{WHITE}{gray}{1}
\definecolor{RED}{rgb}{1,0,0}
\definecolor{GREEN}{rgb}{0,1,0}
\definecolor{BLUE}{rgb}{0,0,1}
\definecolor{CYAN}{cmyk}{1,0,0,0}
\definecolor{MAGENTA}{cmyk}{0,1,0,0}
\definecolor{YELLOW}{cmyk}{0,0,1,0}
}
\definecolor{light-gray}{gray}{0.55}

\usepackage{babel}

\DeclareMathAlphabet{\mathpzc}{OT1}{pzc}{m}{it}
\makeatother

\usepackage{babel}
\begin{document}

\title{Chiral groundstate currents of interacting photons in a synthetic magnetic field}

\author{P. Roushan$^{1}$}
\thanks{These authors contributed equally to this work.}
\author{C. Neill$^{2}$}
\thanks{These authors contributed equally to this work.}
\author{A. Megrant$^{1}$}
\thanks{These authors contributed equally to this work.}
\author{Y. Chen$^{1}$}
\author{R. Babbush$^{3}$}
\author{R. Barends$^{1}$}
\author{B. Campbell$^{2}$}
\author{Z. Chen$^{2}$}
\author{B. Chiaro$^{2}$}
\author{A. Dunsworth$^{2}$}
\author{A. Fowler$^{1}$}
\author{E. Jeffrey$^{1}$}
\author{J. Kelly$^{1}$}
\author{E. Lucero$^{1}$}
\author{J. Mutus$^{1}$}
\author{P. J.J. O'Malley$^{2}$}
\author{M. Neeley$^{1}$}
\author{C. Quintana$^{2}$}
\author{D. Sank$^{1}$}
\author{A. Vainsencher$^{2}$}
\author{J. Wenner$^{2}$}
\author{T. White$^{1}$}
\author{E. Kapit$^{4,5}$}
\author{H. Neven$^{3}$}
\author{J. Martinis$^{1,2}$}

%\email{pedramr@google.com}

\affiliation{$^{1}$Google Inc., Santa Barbara, CA 93117, USA}
\affiliation{$^{2}$Department of Physics, University of California, Santa Barbara, CA 93106, USA}
\affiliation{$^{3}$Google Inc., Los Angeles, CA 90291, USA}
\affiliation{$^{4}$The Graduate Center, CUNY, New York, NY, 10016 USA}
\affiliation{$^{5}$Department of Physics, Tulane University, New Orleans, LA 70118, USA}

\maketitle
%\small
\textbf{The intriguing many-body phases of quantum matter arise from the interplay of particle interactions, spatial symmetries, and external fields\,\cite{Coleman}. Generating these phases in an engineered system could provide deeper insight into their nature and the potential for harnessing their unique properties\,\cite{NoriScience,CiracNP,ColdAtoms,AlanNP,HouckNP,NoriRMP}. However, concurrently bringing together the main ingredients for realizing many-body phenomena in a single experimental platform is a major challenge. Using superconducting qubits, we simultaneously realize synthetic magnetic fields and strong particle interactions, which are among the essential elements for studying quantum magnetism and fractional quantum Hall (FQH) phenomena\,\cite{Tsui1982,Laughlin1983}. The artificial magnetic fields are synthesized by sinusoidally modulating the qubit couplings. In a closed loop formed by the three qubits, we observe the directional circulation of photons, a signature of broken time-reversal symmetry. We demonstrate strong interactions via the creation of photon-vacancies, or "holes", which circulate in the opposite direction. The combination of these key elements results in chiral groundstate currents, the first direct measurement of persistent currents in low-lying eigenstates of strongly interacting bosons. The observation of chiral currents at such a small scale is interesting and suggests that the rich many-body physics could survive to smaller scales. We also motivate the feasibility of creating FQH states with near future superconducting technologies. Our work introduces an experimental platform for engineering quantum phases of strongly interacting photons and highlight a path toward realization of bosonic FQH states.}

It is commonly observed that when the number of particles in a system increases, complex phases can emerge which were absent in the system when it had fewer particles, i.e. the "more is different"\,\cite{Anderson1972}. This observation drives experimental efforts in synthetic quantum systems, where the primary goal is to engineer and utilize these emerging phases. However, it has generally been overlooked that these sought-after phases can only emerge from simultaneous realization and control of particle numbers, real-space arrangements, external fields, particle interactions, state preparation, and quantum measurement. The simultaneous realization of all these ingredients makes synthesizing many-body phases a holistic task, and hence constitutes a major experimental challenge. Engineering these factors, in particular synthesizing magnetic fields, have been performed in several platforms\,\cite{ColdAtoms,Spielman2009,Ketterle2013,Bloch2013,Esslinger2014,HafeziNP2013,Rechtsman2013,LuNpotonics2014,TzuangNP2014,SimonPRX2015}. However, these ingredients have not been jointly realized in any system thus far. To provide a tangible framework, we discuss realization of these key elements in the context of quantum Hall physics, and show when these ingredients come together they can construct a basic building block for creating FQH states.

The FQH states are commonly studied in 2-dimensional electron gases, a fermionic condensed matter system\,\cite{Tsui1982,Laughlin1983}. However, many of the recent advancements in engineered quantum systems are taking place in bosonic platforms\,\cite{NoriRMP,ColdAtoms,CiracNP,AlanNP,HouckNP,NoriScience}. Theoretical studies suggest the existence of rich phases for bosonic FQH systems, similar to their femionic counterparts\,\cite{ZollerSSC,SougatoPRL2008,HaywardPRL,PetrescuPRA2012,IacopoPRL,Hafezi2014}. In particular, bosonic FQH states are known to host non-Abelian anyons, which could implement quantum logic operations through braiding\,\cite{Nayak2008}. Among the prerequisites for realizing bosonic FQH states are (i) strong artificial gauge fields, leading to nearly flat single particle bands, (ii) strong interactions, (iii) low disorder, and (iv) a mechanism for accessing the many-body ground state. In this work, we engineer a modular unit cell consisting of three coupled qubits in a ring, which when tiled can be used to realize FQH phases (Fig.\,1\textbf{(a)} and \textbf{(b)})\,\cite{flatband,Bergholtz2013,Mueller2010}. We concurrently demonstrate tunable gauge fields, strong interactions, and adiabatic groundstate preparation in a low loss and disorder platform, where we have full state preparation and quantum correlation measurement capabilities.

When electrons hop between lattice sites of a crystal placed in a magnetic field, the wavefunction accumulates a path-dependent phase. The interference of electrons traveling along different paths is the fundamental origin of many rich many-body phases seen in correlated systems. However, due to the charge neutrality of photons, they are not affected by physical magnetic fields; therefore, an effective magnetic field has to be synthesized for quantum platforms with bosonic excitations\,\cite{Zoller2003,ColdAtoms,Spielman2009,Hauke2012,Ketterle2013,Bloch2013,Esslinger2014,KochPRA,NunnenkampNJP,ZOltanPRA}. One practical idea, proposed in various settings, suggests that artificial magnetic fields can be created by periodic modulation of the photon hopping strength between the lattice sites\,\cite{Fang2012,EliotUniversal,Hafezi2014}. When the on-site energies of two lattice sites differ by $\Delta$, then sinusoidal modulation of a tunneling term with frequency $\Delta$ and phase $\varphi$ results in an effective complex hopping, where the photon's wavefunction picks up phase $\varphi$ (Fig.\,1\textbf{(c)}). This phase is analogous to the Peierls phase $e\oint \mathbf{A} \cdot d \mathbf{r}$ that is accumulated by a particle of charge $e$ tunneling in an external magnetic vector potential $\mathbf{A}$. This idea can be implemented in a superconducting qubit platform, where qubits play the role of the lattice sites and modulating the strength of the inter-qubit couplings $g$ sets the microwave photon hopping rate.

\begin{figure}
\begin{centering}
\includegraphics[width=87mm]{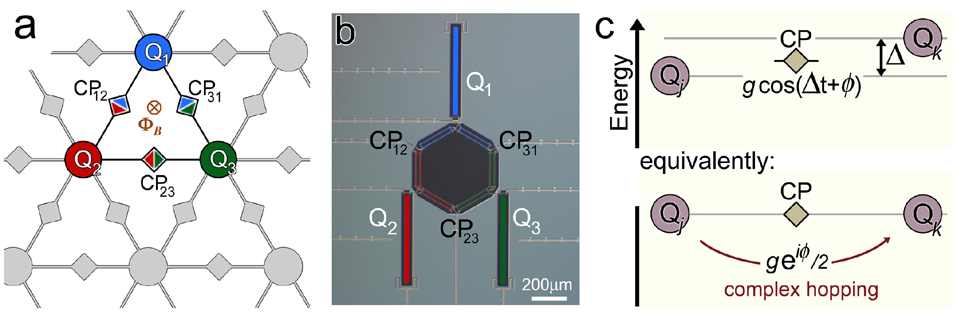} \caption{\textbf{The unit cell for FQH and synthesizing magnetic fields.} \textbf{(a)} A schematic illustration of how qubits and their couplers can be tiled to create a 2D lattice. The 3-qubit unit cell of this lattice, which is realized in this work, is highlighted. \textbf{(b)} An optical image of the superconducting circuit made by standard nano-fabrication techniques. It consist of three superconducting qubits $Q_j$ connected via adjustable couplers $CP_{jk}$. Together, they form a triangular closed loop. \textbf{(c)} A parametric modulation approach is used for synthesizing magnetic fields. If the frequency difference of two qubits is $\Delta$, then the sinusoidal modulation of the coupler connecting them with frequency $\Delta$ and phase $\varphi$ results in an effective resonance hopping ($\Delta=0$) with a complex hopping amplitude between the two qubits.}
\par\end{centering}
\label{fig:chevran}
\end{figure}

\begin{figure}
\begin{centering}
\includegraphics[width=87mm]{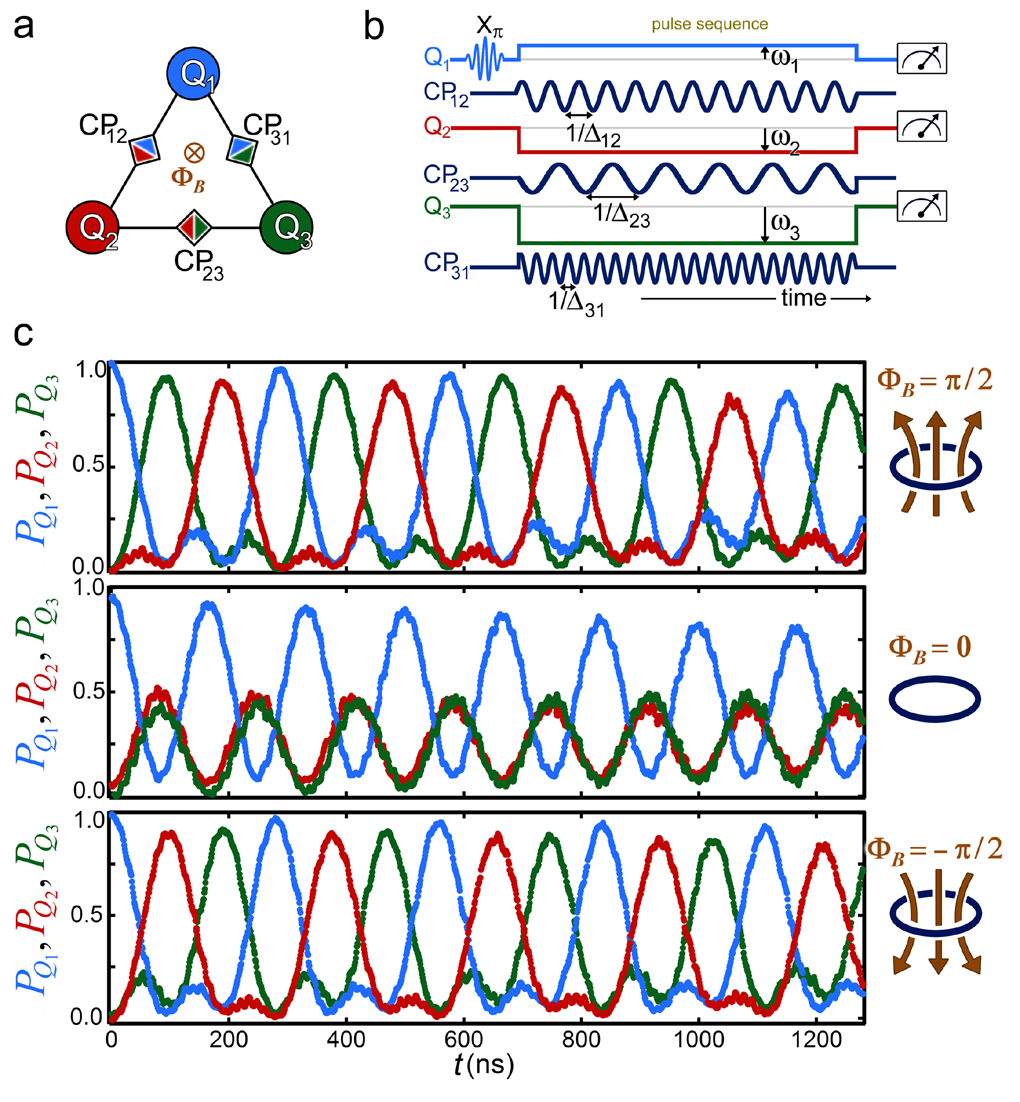} \caption{\textbf{Single-photon circulation resulting from the TRS breaking.} \textbf{(a)} Schematic of the three qubits and their couplers placed in a triangular closed loop.\textbf{(b)} The pulse sequence used for generating and circulating a microwave photon shows that qubits frequencies $\omega_j$ can be chosen to have arbitrary values, but each coupler is needed to modulate with frequency $\Delta_{jk}$, set to the difference in the qubit frequencies that it connects $\omega_j-\omega_k$. The periodic modulation of each coupler can also has a phase $\varphi_{jk}$, where $\Phi_B\equiv\varphi_{12}+\varphi_{23}+\varphi_{31}$. \textbf{(c)} A microwave photon is created by applying a $\pi$-pulse to $Q_1$, at $t=0$ ($\psi_0=|100\rangle$). While applying the pulse sequence shown in \textbf{(b)}, the probability of photon occupying each qubit $P_{Q_{j}}$ as a function of time is measured for three values of $\Phi_B=\pi/2,0,\,-\pi/2$. We use $g_0=4$\,MHz, $\omega_1=5.8$\,GHz, $\omega_2=5.8$\,GHz, $\omega_3=5.835$\,GHz, $\Delta_{12}=0$, $\Delta_{23}=35$\,MHz,$\Delta_{31}=35$\,MHz, $\varphi_{12}=0$, $\varphi_{23}=0$, and $\varphi_{31}$ was used to set $\Phi_B$.}
\par\end{centering}
\label{fig:marching}
\end{figure}

\begin{figure*}
\begin{centering}
\includegraphics[width=142mm]{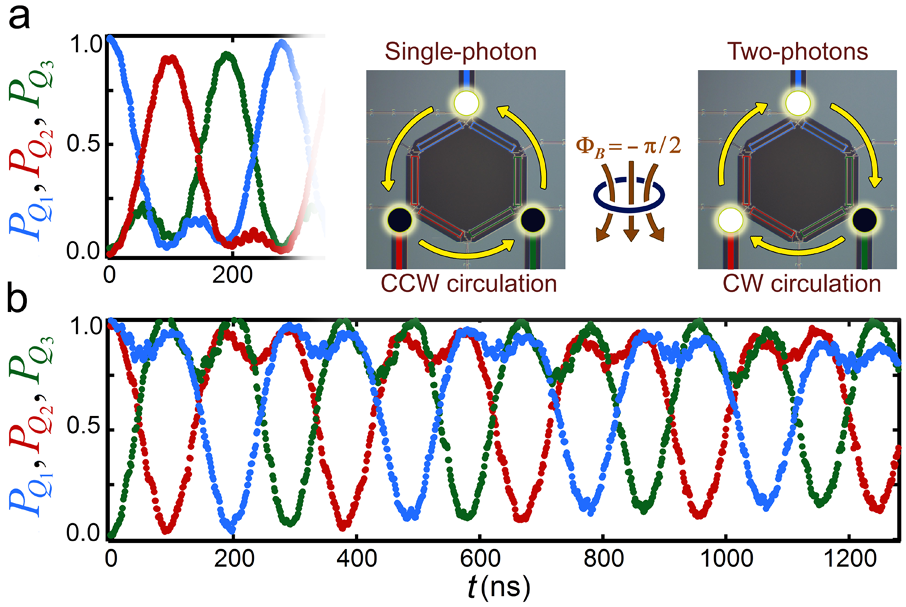}
\caption{\textbf{Signature of strong interaction.} \textbf{(a)} The single-photon circulation data for $\Phi_B=-\pi/2$, which is shown in Fig.\,2\textbf{(c)}, is partially shown for the ease of comparison with the two-photon data shown in \textbf{(b)}. \textbf{(b)} At $t=0$, two photons are created and are occupying $Q_1$ and $Q_2$ sites. They are generated by applying a $\pi$-pulse to $Q_1$ and $Q_2$ and exciting them ($\psi_0=|110\rangle$). The parameters used, pulse sequence, and the measurements are similar to Fig.\,2. While the single-photon circulates in counter-clockwise direction, the photon-vacancy circulates in the clock-wise direction. The counter circulation of the two-photons case compared with the single-photon case is the direct consequence of strong interactions in the system. In the absence of interactions, the direction of circulation would have been the same. These findings are schematically demonstrated in panel \textbf{(a)}. The yellow arrows indicate the direction of circulation of the single-photon or single-vacancy case, where photons and vacancies are depicted by bright and dark disks, respectively, and shown on top of the optical image of the circuit used.}
\par\end{centering}
\label{fig:annealing}
\end{figure*}

We place three transmon superconducting qubits in a ring (Fig.\,1\textbf{(b)}), where each qubit is coupled to its neighbors via an adjustable coupler that can be dynamically modulated on nano-second timescales\,\cite{Yu2014}. The Hamiltonian of the system is
\begin{equation}\label{eq:fullH}
H(t)=\hbar\sum\limits_{j=1}^3 \omega_j (\hat{n}+1/2)+\hbar\sum\limits_{j,k} g_{jk}(t)(a_{j}^{\dagger}a_{k}+a_{j}a_{k}^{\dagger})+H_{\mbox{int}},
\end{equation}
where $a^{\dagger} (a)$ are bosonic creation (annihilation) operators, $\omega_j$ is frequency of qubit $Q_j$, $\hat{n}=a_{j}^{\dagger}a_{j}$ is the particle number operator, and $g_{jk}$ is the strength of the inter-qubit coupling between qubits $Q_j$ and $Q_k$. $H_{\mbox{int}}$ captures the interaction between bosons and is set by the non-linearity of the qubits. This term does not affect the dynamics in the single-photon manifold, and we will discuss its role in the two-photon manifold in more detail later. We modulate $g$ of each coupler according to $g_{jk}(t)=g_0 \cos(\Delta_{jk}t+\varphi_{jk})$, and choose $\Delta_{jk}$ to be the difference between the frequencies of the two qubits that the coupler connects, i.e. $\Delta_{jk}=\omega_j-\omega_k$ \,(Fig.\,2\textbf{(b)}). If $|g_{jk}|<<|\omega_j-\omega_k|$, then, in the rotating frame, the effective Hamiltonian of the system becomes

\begin{equation}\label{eq:RWAH}
H_{\mbox{eff}}(\Phi_B)=\frac{\hbar}{2}\sum\limits_{j,k} g_{0}(e^{i\varphi_{jk}}a^{\dagger}_{j}a_{k}+e^{-i\varphi_{jk}}a_{j}a^{\dagger}_{k}),
\end{equation}
where $\Phi_B\equiv\varphi_{12}+\varphi_{23}+\varphi_{31}$ is the effective magnetic flux and is gauge-invariant\,\cite{supp}. One can intuitively understand the origin of the gauge invariance of $\Phi_B$ by noting that the three qubits in our case form a closed loop, and the accumulated phase needs to be single-valued when going around this loop. In other words, if the qubits' loop were open, $\Phi_B$ would not be gauge-invariant (see \,\cite{supp} for details).

Based on this idea, we construct a protocol (Fig.\,2\textbf{(b)}) and study the dynamics of single microwave photons in our system. At $t=0$, we create a microwave-photon which occupies $Q_1$ ($\psi_{0}=|100\rangle$), and measure $P_{Q_j}$, the photon occupation probability of $Q_j$, as a function of time. As shown in the middle panel of Fig.\,2\textbf{(c)}, the photon has a symmetric evolution for $\Phi_B=0$. It propagates from $Q_1$ to $Q_3$ and $Q_2$ simultaneously, then back to $Q_1$, and then repeats the pattern with no indication of any preferred circulation direction $\def\arraystretch{0.65}(\mbox{blue}\rightarrow\begin{array}{c}\mbox{red}\\\mbox{green}\end{array}\rightarrow\mbox{blue}\rightarrow...)$. Setting $\Phi_B=\pi/2$ leads to fundamentally different dynamics, where the photon propogation shows a preferred circulation direction and marches in a clockwise order from $Q_1$, to $Q_3$, to $Q_2$, eventually back to $Q_1$, and then repeating the pattern $(\mbox{blue}\rightarrow\mbox{green}\rightarrow\mbox{red}\rightarrow\mbox{blue}\rightarrow...)$. Choosing $\Phi_B=-\pi/2$ leads to counter-clockwise circulation, demonstrating that the synthetic flux $\Phi_B$ behaves quite similarly to physical magnetic flux\,\cite{supp}.

The hallmark of magnetic fields in a system is the breaking of time reversal symmetry (TRS). Commonly, TRS preserving evolution of the state is defined as $\psi(t)=\psi(-t)$. Verifying TRS breaking based on this relation in a real experiment can be difficult, since reversing the flow of time is generally not feasible. However, the dynamics considered here is periodic with period $T=280$\,ns for $\Phi_B=\pm\pi/2$ and $T=170$\,ns for $\Phi_B=0$ case. This periodicity allows us to arrive at a practical definition for TRS, which is $\psi(t)=\psi(T-t)$; e.g., one could follow the evolution of state from $t=T$ backward and see if it is the same as going forward from $t=0$. It can be seen in Fig.\,2\textbf{(c)} that TRS is preserved for $\Phi_B=0$ and is broken when $\Phi_B=\pm\pi/2$. These observations establish TRS breaking for $\Phi_B=\pm\pi/2$ and further illustrate that the synthetic flux $\Phi_B$ indeed behaves akin to physical magnetic flux. The quantum nature of the circulation is manifested through quantum correlation measurements which show entanglement between qubits(see \,\cite{supp} for data). The measured entanglement makes our experiment distinct from others which are based on classical wave mechanics or those where the time-scales are much longer than the quantum coherence of the system, i.e. are in semi-classical limit \,\cite{Fang2012,Fang2013,Estep2014,Lehnert2015,HafeziNP2013,Raghu2008, Wang2009,Rechtsman2013,Khanikaev2013, LuNpotonics2014,TzuangNP2014,SimonPRX2015}.

\begin{figure*}
\begin{centering}
\includegraphics[width=160mm]{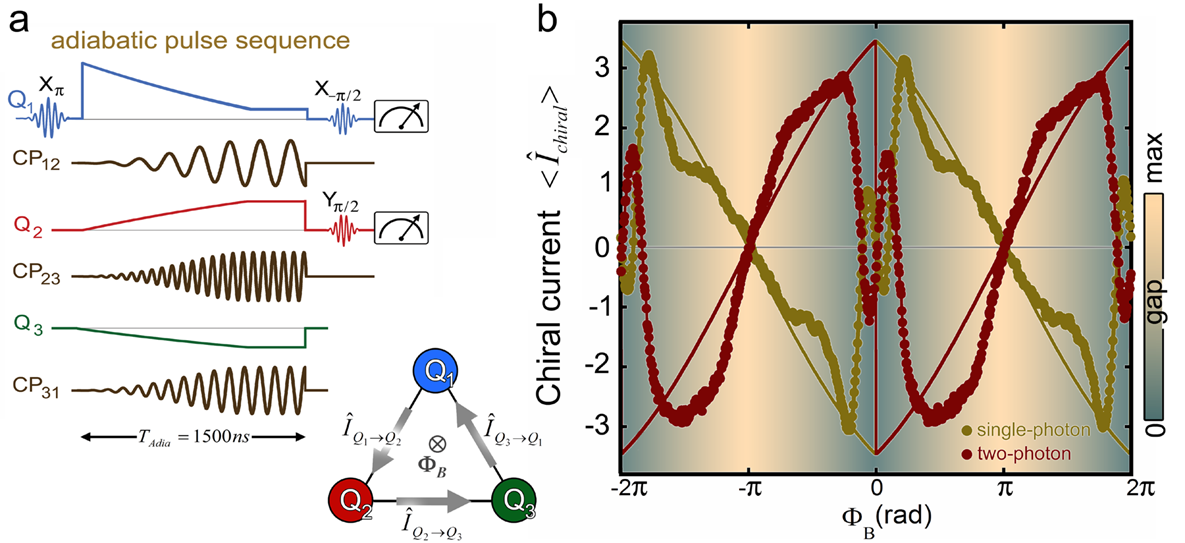}
\caption{\textbf{Chiral currents in the groundstate.} \textbf{(a)} The pulse sequence for adiabatically preparing the groundstate of Eq.(2). For groundstates in the single-photon manifold, $Q_1$ is excited at $t=0$ ($\psi_0=|100\rangle$), and in the two-photon manifold, $Q_1$ and $Q_2$ are excited ($\psi_0=|110\rangle$). To measure $\hat{I}_{Q_1\rightarrow Q_2}$ at the end of parameter ramping, $Q_1$ and $Q_2$ are rotated, allowing for measurements of $\langle\sigma^X_{Q1}\sigma^Y_{Q2}\rangle$ and $\langle\sigma^Y_{Q1}\sigma^X_{Q2}\rangle$ (see \,\cite{supp} for details). \textbf{(b)} The measured values of $\langle\hat{I}_{\mbox{chiral}}\rangle$ in the single-photon (olive-color) or two-photon manifolds (maroon-color). The solid lines are computations for $T_{\mbox{adia}}\rightarrow\infty$. The energy gap of the Hamiltonian of the system (Eq.(2)) as a function of $\Phi_B$ is numerically computed and is shown as the background of the data. The gap closes at $\Phi_B=0,\pm2\pi$ and the groundstate become degenerate (green regions). The maximum gap size is $3g_0$, which here is $12$\,MHz.}
\par\end{centering}
\label{fig:annealing}
\end{figure*}

We next focus on signatures of strong interactions, which are vital for realizing FQH states, as the many-body gap is set by the smaller of $g$ and $U$. The typical weakness of interactions between bosons makes studying many-body quantum phenomena a major engineering challenge\,\cite{ZollerSSC}. Superconducting qubits, however, naturally overcome this challenge and provide a platform where microwave photons can have strong interactions. Systems of coupled qubits can be understood with a Bose-Hubbard model, where the on-site interaction $U$ originates from the expansion of qubit's confining cosine potential:

\begin{equation}\label{eq:Hint}
H_{\mbox{int}}=- \frac{U_2}{2} \sum_{j} \hat{n}_j (\hat{n}_j - 1) + \frac{U_3}{6} \sum_j \hat{n}_j (\hat{n}_j - 1) (\hat{n}_j - 2) + ... .
\end{equation}
In our system $U_2\approx U_3\sim200$\,MHz which sets the energy difference between single and double photon occupancy; e.g. the $|200\rangle$ to $|110\rangle$ transition. The hopping "bandwidth" in each manifold is set by $g$ and is a few MHz. Therefore $U\gg g$ and qubits effectively form a hard core boson system.

The signature of strong interactions can be seen in the two-photon circulation as shown in Fig.\,3\textbf{(b)}. In the absence of interactions one expects that two photons to circulate freely with the same chirality as a single photon. However, two-photon circulation in our system displays the opposite chirality, indicating that, as a result of strong interactions, photons do not move freely. Consequently, given that our system has three sites, when two photons are injected it is more natural to consider the motion of the photon-vacancy. Similar to the physics of holes in an electron band, the photon-vacancies have the opposite "charge", and hence circulate in the opposite direction compared to photons.

In condensed matter systems, one is generally interested in finding the groundstate of a many-body system and probing its properties. In particular, the key signature of FQH states is the appearance of groundstate chiral edge currents. As the many-body Chern number of an FQH phase can be extracted from the DC conductivity tensor, the capability to measure ground state currents is especially valuable. Although the evolution of $|100\rangle$ or $|110\rangle$, as discussed so far, provides an intuitive understanding of the response of the system to this synthetic gauge, these data do not directly reflect the groundstate properties of the system, because these initial states are not eigenstates of the Hamiltonian. To study ground state properties, we adiabatically prepare groundstates of Eq.(2) and examine breaking the TRS by measuring the chiral current in the groundstates (see Fig.\,4\textbf{(a)} for pulse sequence). Analogous to the continuity equation in classical systems, a current operator $\hat{I}$ can be defined by equating the current in and out of a qubit site to the change of the photon number operator on that site ($\hat{I}_{in}-\hat{I}_{out}=d\hat{n}/dt$) \,\cite{supp}. From the continuity equations, we define the chiral current operator to be

\begin{equation}\label{eq:Icirc}
\hat{I}_{\mbox{chiral}}\equiv\sum\limits_{j,k}\hat{I}_{Q_j\rightarrow Q_k}=i\sum\limits_{j,k} (e^{i\varphi_{jk}}a^{\dagger}_{j}a_{k}-e^{-i\varphi_{jk}}a_{j}a^{\dagger}_{k}).
\end{equation}
Since $\hat{I}_{\mbox{chiral}}$ flips sign under TRS, we expect that its ground state expectation value will be zero whenever the Hamiltonian is TRS preserving, and nonzero otherwise. This equilibrium current is distinct from the commonly measured non-equilibrium particle imbalance\,\cite{Imbalance2014,MoreImbalance2014,BlochChiralCurrent}, as experimental measurement of $\hat{I}_{\mbox{chiral}}$ require access to the groundstate.

To measure $\langle\hat{I}_{\mbox{chiral}}\rangle$ in the single-photon manifold, initially we prepare $\psi_{0}=|100\rangle$ followed by a ramp up of the Hamiltonian parameters to generate Eq.(2) for various $\Phi_B$ values (olive color, Fig.\,4\textbf{(b)}). For preparing groundstates in the two-photon manifold, we initially create $\psi_{0}=|110\rangle$ by exciting two qubits, followed by a similar ramp and measurements (maroon color, Fig.\,4\textbf{(b)}). Note that due to large $U/g$ ratio, the two-photon manifold with and without double occupancies are almost entirely separate. Because of the three-fold symmetry of the system, measuring the current operator between any pair of qubits, e.g. $\hat{I}_{Q_1\rightarrow Q_2}$, suffices for knowing $\langle\hat{I}_{\mbox{chiral}}\rangle$. The solid lines are from numerical computations assuming perfect adiabaticity. For a given $\Phi_B$, the measured $\langle\hat{I}_{\mbox{chiral}}\rangle$ on single- and two-photon manifold show almost exactly opposite values indicating that photons and photon-vacancies have opposite chiralities. On both manifolds and away from the origin, $\langle\hat{I}_{\mbox{chiral}}\rangle$ rather abruptly become non-zero with opposite values for $\Phi_B>0$ and $\Phi_B<0$, showing a quantum transition.  Additional interesting points are $\Phi_B=\pm\pi$, where $\langle\hat{I}_{\mbox{chiral}}\rangle$ goes to zero on both one-photon and two-photon manifolds, and in contrast to $\Phi_B=0$, the measured chiral current close to $\Phi_B=\pm\pi$ is smooth.

The vanishing of $\langle\hat{I}_{\mbox{chiral}}\rangle$ at $\Phi_B= 0,\pm\pi$ can be understood by noticing that the Hamiltonian of the system is real at these points and hence cannot break the TRS, whereas for other values it is irreducibly complex. Several feature of the data can be understood by computing the gap between the groundstate and the first excited state(background color of Fig.\,4\textbf{(b)}). For $\Phi_B = 0$, $\langle\hat{I}_{\mbox{chiral}}\rangle$ is discontinuous, as the ground state is degenerate at $\Phi_B = 0$ and any finite $\Phi_B$ breaks this degeneracy and leads to chiral currents, effectively producing a first-order phase transition. On the other hand, for $\Phi_B=\pm\pi$, the ground state is not degenerate and there is a large gap to the excited states, and $\langle\hat{I}_{\mbox{chiral}}\rangle$ must therefore smoothly cross zero as $\Phi_B$ crosses $\pm\pi$. The origin of the oscillatory behavior close to $\Phi_B=0$ is also due to gap closing, as a result of which the adiabatic ramps become incapable of providing correct results.

Our experiment highlights the strengths of superconducting qubits for synthesizing many-body phases of quantum matter. The inherent simplicity of the coupling modulation method also played a key role in this first demonstration of synthetic gauge fields with superconducting qubits; frequently, synthetic gauge field proposals for superconducting circuits demand challenging new architectures and are susceptible to charge noise. The scheme we employed avoids these issues, can be generally applied for other applications\,\cite{EliotUniversal}, and highlights a path forward\,\cite{supp} beyond these proof of principle experiments to the direct realization of FQH states. To realize FQH physics, the system must be large compared to the magnetic length $l_B$ of the Hamiltonian. If we choose the Kapit-Mueller Hamiltonian\,\cite{Mueller2010} as a basis, a flux per plaquette $\Phi_B=1/3$ yields $l_B = 0.69$, which suggests an $L \times L$ lattice with $L \geq 6$ as an appropriate host for FQH physics. Further, a $2 \times L$ ladder with nearest and next nearest neighbor hopping can host a nearly exact Laughlin ground state that displays many of the properties of its $L \times L$ parent state. These include a local excitation gap, fractionalized excitations and a topological degeneracy which manifests as charge density wave order in ladder systems\,\cite{Flavin2011}. For both host systems, the Laughlin ground state is resilient against local phase noise, and it can be prepared through adiabatic evolution or resonant sequential photon injection, or stabilized indefinitely through engineered dissipation\,\cite{EliotPRX2014,HakanPRL}. Thus, simply increasing the size of our system provides a near-term experimental path for generating FQH states of light.

\footnotesize \textbf{Acknowledgments:} We acknowledge discussions with L. Lamata, A. Rahmani, E. Rico, M. Sanz and E. Solano. Devices were made at the UCSB Nanofab Facility, part of the NSF-funded NNIN, and the NanoStructures Cleanroom Facility.

\footnotesize \textbf{Correspondence:} All correspondence should be addressed to pedramr@google.com.

\footnotesize \textbf{Author Contributions:} P.R., C.N., and A.M. performed the experiment. E.K. provided theoretical assistance. P.R. analysed the data, and with C.N. and E.K.  co-wrote the manuscript and supplementary information. All of the UCSB and Google team members contributed to the experimental setup. All authors contributed to the manuscript preparation.


\begin{thebibliography}{50}
\bibitem [{fla()}]{flatband}%
  \BibitemOpen
  \href@noop {} {}\bibinfo {note} {At low (uniform) flux densities, the single
  particle band is nearly flat in this triangular configuration. However, we
  note that to obtain a denser FQH states hopping beyond nearest neighbors is
  generally required to ensure band flatness.}\BibitemShut {Stop}%
\end{thebibliography}
\end{document}

\title{Supplementary materials:
Chiral groundstate currents of interacting photons in a synthetic magnetic field}
\author{P. Roushan$^{1}$}
\thanks{These authors contributed equally to this work.}
\author{C. Neill$^{2}$}
\thanks{These authors contributed equally to this work.}
\author{A. Megrant$^{1}$}
\thanks{These authors contributed equally to this work.}
\author{Y. Chen$^{1}$}
\author{R. Babbush$^{3}$}
\author{R. Barends$^{1}$}
\author{B. Campbell$^{2}$}
\author{Z. Chen$^{2}$}
\author{B. Chiaro$^{2}$}
\author{A. Dunsworth$^{2}$}
\author{A. Fowler$^{1}$}
\author{E. Jeffrey$^{1}$}
\author{J. Kelly$^{1}$}
\author{E. Lucero$^{1}$}
\author{J. Mutus$^{1}$}
\author{P. J.J. O'Malley$^{2}$}
\author{M. Neeley$^{1}$}
\author{C. Quintana$^{2}$}
\author{D. Sank$^{1}$}
\author{A. Vainsencher$^{2}$}
\author{J. Wenner$^{2}$}
\author{T. White$^{1}$}
\author{E. Kapit$^{4,5}$}
\author{J. Martinis$^{1,2}$}

%\email{pedramr@google.com}

\affiliation{$^{1}$Google Inc., Santa Barbara, CA 93117, USA}
\affiliation{$^{2}$Department of Physics, University of California, Santa Barbara, CA 93106, USA}
\affiliation{$^{3}$Google Inc., Los Angeles, CA 90291, USA}
\affiliation{$^{4}$The Graduate Center, CUNY, New York, NY, 10016 USA}
\affiliation{$^{5}$Department of Physics, Tulane University, New Orleans, LA 70118, USA}

\let\oldthebibliography=\thebibliography
\let\oldendthebibliography=\endthebibliography
\renewenvironment{thebibliography}[1]{%
    \oldthebibliography{#1}%
    \setcounter{enumiv}{5}%
}{\oldendthebibliography}

\maketitle
\textbf{}
%\tableofcontents
%\clearpage

\section{\textcolor{SECTIONCOLOR}{1. Device: the superconducting qubits with gmon architecture}}

\begin{figure*}
\begin{centering}
\includegraphics[width=175mm]{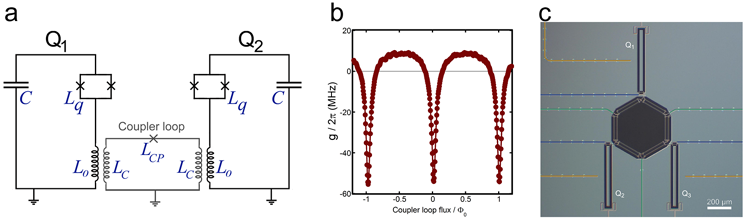}
\par\end{centering}
\caption{ \textbf{Device architecture.} \textbf{(a)} The circuit diagram of two superconducting qubits connected with an adjustable coupler. Each qubit is a non-linear LC resonator, and the two qubits are inductively coupled to a coupler loop through the mutual inductance between their $L_0$ and the loop's $L_C$. The coupler loop has a single Josephson junction with inductance $L_{CP}$, which can be tuned by applying magnetic flux into the coupler loop, allowing variable coupling strength between the two qubits $g$. \textbf{(b)} The measured value of $g$ as a function of flux into the coupler loop. \textbf{(c.)} An optical micrograph of the three qubit device used in this work with coupling between each pair of them.  Gray regions correspond to aluminum; black regions are where the aluminum has been etched away to expose the underlying sapphire substrate to define the qubits and wiring. Microwave drive lines which are used to excite qubits are shown in light brown. The flux bias lines, which are used to bring qubit on and off resonance, are highlighted with blue, and the lines used to adjust flux into coupling loops are highlighted with green.}
\label{fig:device}
\end{figure*}

In this section, we briefly discuss the working principle of the coupled superconducting qubits used in this work. For a detailed discussion, please see references\,\cite{Geller2015,Yu2014}.
\vspace{2mm}

Our superconducting qubits are non-linear LC resonators composed of a capacitor $C$, a DC SQUID with total inductance of $L_q$, and an inductor $L_0$ in series with $L_q$ to ground. The capacitance and SQUID form the basis of the standard Xmon qubit \,\cite{RamiPRL} with the added inductor allowing for tunable coupling to a neighboring qubit. A circuit diagram of two coupled qubits is shown schematically in Fig.~\ref{fig:device}\textbf{(a)}. We couple qubits with an inductive coupler loop, which allows changing the strength of the qubit-qubit interaction $g$, hence the name "gmon" \,\cite{Geller2015,Yu2014}.

The adjustable coupling in gmon qubits can be intuitively understood by comparing them with conventional variometers. Variometers are transformers capable of varying the mutual inductance between their primary and secondary solenoid coils by changing the angle between the axis of these two coaxial inductors. In the gmon architecture, the same functionality is achieved by the coupler loop. An excitation in either qubit generates a current in this loop which then excites the neighboring qubit. Changing the magnetic flux through the coupler loop is analogous to rotating the axis of the solenoids in a variometer, and allows tuning the coupling between the two qubit loops. This is because the magnetic flux sets the effective junction impedance of $L_{CP}$. If the inductance is large, then a smaller current will flow through the coupler loop and the coupling become weaker. This qubit design enables us to continuously vary the coupling strength $g$ over nanosecond timescales without any degradation in the coherence of the qubits\,\cite{Geller2015,Yu2014}. As shown in Fig.~\ref{fig:device}\textbf{(b)}, $g/2\pi$ can take any value between $-55$ MHz and $+5$ MHz, including zero.

In this work, we placed three qubits in a triangular loop and implemented an adjustable coupling between every pair of qubits (see Fig.~\ref{fig:device}\textbf{(c)}). Synthesizing gauge fields requires periodically modulating the flux into the three coupling loops, with various frequencies on the order of tens of MHz. This requirement leads to an arbitrary pulse sequence (see Fig. 2 of the main text) that need careful calibration to allow observation of the patterns such as those shown in Fig.2 of the main text. For a detailed discussion of calibration routines, see references \cite{Yu2014,Ergodicity2015}.

The device was fabricated using standard optical and e-beam lithography techniques, discussed in \,\cite{RamiPRL} and is benefited from the low-loss crossovers discussed in\,\cite{JimmyAirBridge}. The qubit frequencies are tunable, but mainly flux biased to around $6$ GHz, with non-linearities close to $210$ MHz. The energy relaxation time, \(T_1\), is \(\sim10\mu s\), and the de-coherence time, \(T_2\) is \(\sim2\mu s\). The experiment was performed at the base temperature of a dilution refrigerator ($\sim$20 mK).
%\clearpage

\section{\textcolor{SECTIONCOLOR}{2. Method: synthesizing gauge fields with AC modulation of inter-qubit couplings}}

In this section, we discuss the theory implemented for realizing complex hopping terms in our superconducting qubit system and present the logic behind the equations used in the main text. For a detailed discussions, please see references \cite{EliotPRA,EliotUniversal}. Also, we provide the rationale for the current operator defined and show intuitively why this quantity provide a measure of chirality.

For quantum particles hopping on a lattice, an external gauge field $\mathbf{A}$ causes the tunneling terms between nearby sites to become complex, with the Peierls tunneling phase accumulated when tunneling between sites $j$ and $k$ given by $\varphi_{jk} \equiv e \int_{r_{j}}^{k} \mathbf{A} \cdot d \mathbf{r}$. This modifies the tunneling term $g_{jk} \of{a_j^{\dagger} a_{k} + a_{k}^{\dagger} a_{j} } \to g_{jk} \of{a_j^{\dagger} a_{k} e^{i \phi_{jk}} + a_{k}^{\dagger} a_{j} e^{-i \phi_{jk}} }$, and breaks time reversal symmetry, as the time reversal symmetry operator $T$ is antiunitary and enacts charge conjugation. The complex phase in $H$ between any two sites can be eliminated through a local unitary transformation $\ket{\psi} \to e^{i \of{\alpha n_{j} + \beta n_{k} } } \ket{\psi}$ (equivalent to shifting $\mathbf{A}$ by the gradient of a scalar function), but the sum of the phases $\varphi_{jk}$ along any closed loop is a gauge invariant quantity that is invariant under any local unitary transformations. So long as this phase is nonzero modulo $2 \pi$, the effective magnetic flux $\Phi_B$ through the loop is nonzero, with real physical consequences for the system's time evolution.

To engineer these phases in a qubit array, it is sufficient to consider a pair of qubits coupled by a real, time dependent exchange coupling $g \of{t}$. We let the energy of qubit 1 be equal to $\omega$ and the energy of qubit 2 be equal to $\omega + \Delta$. Our two-qubit Hamiltonian becomes
\begin{eqnarray}
H = \omega n_1 + \of{\omega + \Delta} n_2 + g \of{t} \of{ a_{1}^{\dagger} a_2 + a_{2}^{\dagger} a_1}.
\end{eqnarray}
If we assume that $\abs{g \of{t}} \ll \Delta$ and initialize the system with a single photon in one of the two qubits, then the photon will remain at that qubit indefinitely, as the two qubits are far off-resonant from each other. To exchange photons between the qubits, we must oscillate $g \of{t}$, i.e.
\begin{eqnarray}
g \of{t} = 2 g \cos \of{ \Delta t + \varphi}, %\\
%H = \omega n_1 + \of{\omega + \Delta} n_2 + g \of{e^{i \Delta t + i \varphi} + e^{-i \Delta t - i \varphi}} \of{ a_{1}^+ a_2 + a_{2}^+ a_1}. \nonumber
\end{eqnarray}
\begin{eqnarray}
%g \of{t} = 2 g \cos \of{ \Delta t + \varphi}                                  , \\
H = \omega n_1 + \of{\omega + \Delta} n_2 + g \of{e^{i \Delta t + i \varphi} + e^{-i \Delta t - i \varphi}} \of{ a_{1}^{\dagger} a_2 + a_{2}^{\dagger} a_1}. \nonumber
\end{eqnarray}

We now move to the rotating frame via the unitary transformation $\ket{\psi} \to e^{i \Delta n_2 t} \ket{\psi}$. Incorporating this transformation into the time dependent Schrodinger equation $i \partial_t \ket{\psi} = H \of{t} \ket{\psi}$ we get
\begin{eqnarray}
H = \omega \of{ n_1 + n_2 } + g \of{e^{i \Delta t + i \varphi} + e^{-i \Delta t - i \varphi}} \of{ a_{1}^{\dagger} a_2 e^{-i \Delta t} + a_{2}^{\dagger} a_1 e^{i \Delta t}}.
\end{eqnarray}
Expanding the tunnel coupling leaves a pair of terms which are time-independent and a pair terms which rapidly oscillate at $\pm 2 \Delta t$, which we can ignore in the rotating wave approximation, valid if we assume $\abs{g} \ll \Delta$. We are thus left with a final Hamiltonian which is time independent but complex,
\begin{eqnarray}
H = \omega \of{ n_1 + n_2 } + g \of{ a_{1}^{\dagger} a_2 e^{-i \varphi} + a_{2}^{\dagger} a_1 e^{i \varphi}}.
\end{eqnarray}
Thus, just as charge conservation leads to nontrivial phases for electrons moving in a real magnetic field, energy conservation leads to nontrivial phases for tunneling photons, since the photon must accumulate the phase of the drive field when it gains or loses energy to tunnel between qubits. Since this example only considers two sites, we can eliminate this phase through a unitary transformation (equivalent to choosing a different origin of time $t=0$), but when we consider a closed loop of three or more qubits with drive fields that differ in phase by values other than $0$ or $\pi$, we can no longer regain time reversal symmetry by choosing an appropriate origin for time so that $H\of{t} = H\of{-t}$, and are thus left with a nontrivial artificial magnetic flux in our rotating frame Hamiltonian.

To define a current operator, we consider the continuity equation for each site $j$, $\partial_{t} n_{j} = I_{in} - I_{out}$. Since $\partial_{t} n_{j} = -i \sqof{H, n_{j}}$, we have three equations for qubits 1,2,3:
\begin{eqnarray}
\partial_{t} n_{j} = -i \sqof{H,n_{j}} = - i \of{a_{j} a_{j-1}^{\dagger} e^{-i \varphi_{jj-1}} - a_{j}^{\dagger} a_{j+1}  e^{-i \varphi_{jj+1}} } \equiv \of{ - I_{jj-1} + I_{jj+1} }.
\end{eqnarray}
From these equations, we readily define the current $I_{jk}$ between qubits $j$ and $k$ to be:
\begin{eqnarray}
I_{jk} = i \of{a_{j}^{\dagger} a_k e^{i \varphi_{jk}} - a_{j} a_k^{\dagger} e^{-i \varphi_{jk}} }.
\end{eqnarray}
Since our Hamiltonian is uniform and has magnetic translational symmetry, to measure the current in any eigenstate it is sufficient to measure the current through a single link.

\section{\textcolor{SECTIONCOLOR}{3.Supplementary data}}

In this section, we provide additional experimental data to better explain the method we used for realizing complex hopping, and to provide a deeper insight into the physics of the excitation circulation among the three qubits. At the end we outline the next stages of this project and provide a road-map for realization of FQH states with superconducting qubits.

\begin{figure}
\begin{centering}
\includegraphics[width=153mm]{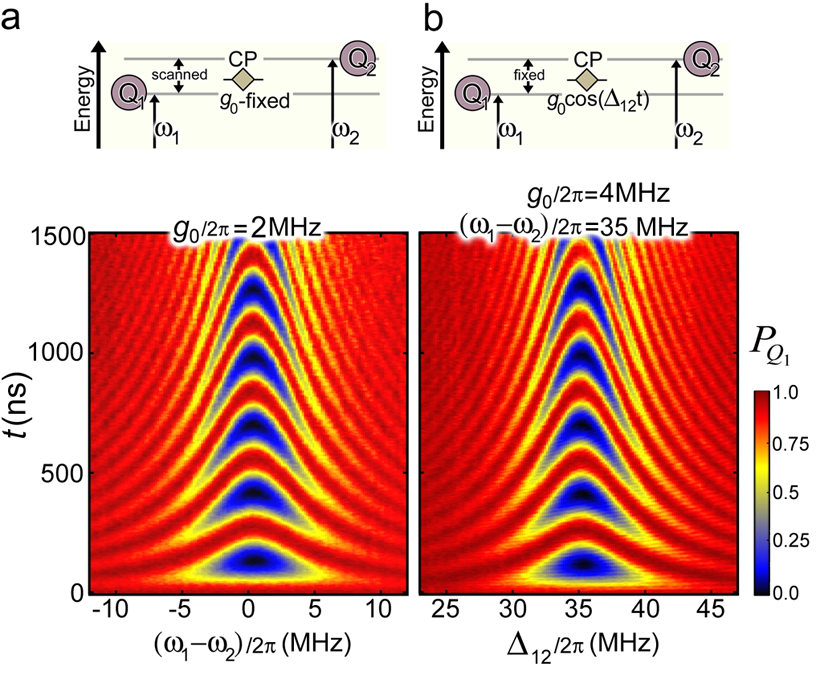} \caption{\textbf{On and off resonance tunneling.} Two qubits $Q_1$ and $Q_2$ with $|0\rangle\rightarrow|1\rangle$ transitions of $\omega_{1}$ and $\omega_{2}$, respectively, are connected via a coupler of strength $g$. At $t=0$, $Q_1$ is excited and its photon occupation probability, $P_{Q_1}$, is measured as a function of time.  \textbf{(a) Fixed coupling}. The frequency difference of the two qubits $(\omega_{1}-\omega_{2})/2\pi$ is varied, while coupling is fixed at $g_0/2\pi=2$\,MHz. \textbf{(b) Periodically modulating coupling}. $Q_1$ and $Q_2$ are set to a fixed detuning $(\omega_{1}-\omega_{2})/2\pi=35$\,MHz and the coupling frequency $\Delta_{12}$ is varied while its amplitude is fixed to $g_0/2\pi=4$\,MHz. The measured chevron patterns are nominally identical.}
\par\end{centering}
\label{fig:chevran}
\end{figure}

\vspace{6mm}
\psection{Parametric modulation of hopping.} The basic idea of parametric modulation of the hopping term can be implemented in a superconducting qubit platform, where qubits play the role of the lattice sites and modulating the strength of the inter-qubit couplings $g$ sets the microwave photon hopping rate (Fig. S2). Hopping also depends on the on-site energies $\omega$, and Fig. S2 demonstrates its interplay with $g$. In a system of two coupled qubits $Q_1$ and $Q_2$, at $t=0$ we excite $Q_1$ and measure its photon occupation probability,$P_{Q_1}$, as a function of time $t$ and on-site energy differences $\omega_1-\omega_2$. For a constant $g$, when on-site energy differences are larger than $g$, the hopping is impeded (away from the center in panel\,\textbf{(a)}). However, if $g$ is modulated with the frequency of the on-site energy difference of the sites that it connects, then photon hopping would be restored (panel\,\textbf{(b)}). In spite of the astonishing similarity of the two data sets, the hopping in \textbf{(b)} is not generally equivalent to \textbf{(a)}, and has the major advantage that its control sequence can be utilized for synthesizing magnetic fields. The key idea is that in \textbf{(b)} the photon's wavefunction can pick up the phase of the modulation during hopping. This phase is analogous to the Peierls phase $ e \oint_{r_{i}}^{r_{j}} \mathbf{A} \cdot d \mathbf{r}$ accumulated by a particle of charge $e$ tunneling in an external magnetic vector potential $\mathbf{A}$.

\begin{figure*}
\begin{centering}
\includegraphics[width=169mm]{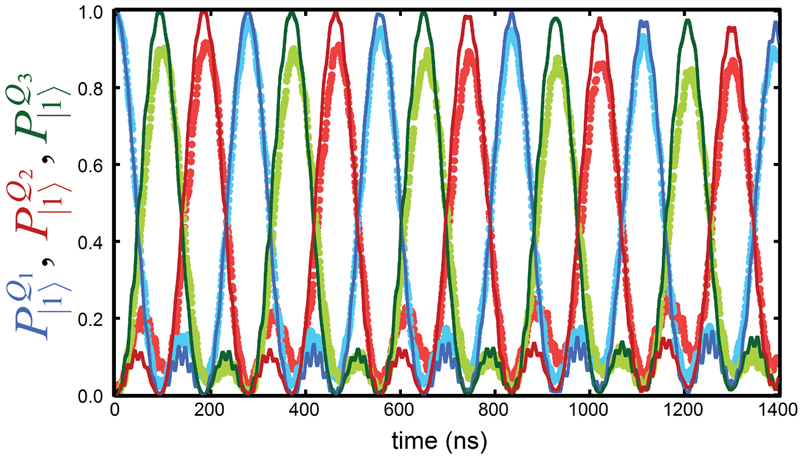}
\par\end{centering}
\caption{ \textbf{Numerical computation of the circulation patterns}. Using Eqn.\,(2) of the main text, the measured evolution of $|100\rangle$ is fitted with a single fitting parameter $g_0/2\pi=$\,4.1MHz. The experimentally measured data points are shown in brighter colors and the fittings are presented with solid, thin darker lines. The fitting does not consider any decaying or decoherence mechanism. For the experiment, we use $\omega_1=5.8$\,GHz, $\omega_2=5.8$\,GHz, $\omega_3=5.835$\,GHz, $\Delta_{12}=0$, $\Delta_{23}=35$\,MHz, $\Delta_{31}=35$\,MHz, $\varphi_{12}=0$, $\varphi_{23}=0$, and $\varphi_{31}$ was used to set $\Phi_B$.  }
\label{fig:fit}
\end{figure*}

\vspace{3mm}

\psection{Decoherence effects.} Given the small size of our system, one can numerically generate the measured circulation patterns and fit the data. We adopted the time dependent Hamiltonian given in Eq. (1) of the main text and used only one fitting parameter, which is $g_0/2\pi=$\,4.1MHz. Our fit shows a remarkably good agreement with data. The fast ripples, seen close to zero excitation in the data, are also observed in the fits, indicating that they originate from unitary, counter-rotating corrections to the rotating frame Hamiltonian and not incoherent processes or experimental errors. Remarkably, the effect of the two dominant error channels ($T_{1}$ photon losses and $T_{2}$ dephasing) is negligible over the window of time plotted. For photon losses, this is simply because the average lifetime $T_{1} \sim 10 \mu s$ is much larger than the duration of the experiment, so photon losses are rare. The absence of phase noise in the plot, however, is a more subtle point, and stems from the basic fact that a single number $T_{2}$ does not capture all of the physics of dephasing processes. Unlike photon losses (which have a noise power spectrum that is approximately flat in our regime of operation), phase noise is generated by random $1/f$ and telegraph fluctuators\,\cite{PeterNoise}, which have a power spectrum that is peaked at $\omega = 0$ and decays to zero as $\omega$ becomes large. This low-frequency peak has dramatic consequences for free Ramsey  decay (where there is no Hamiltonian that anticommutes with the fluctuating $\delta \of{t} \sigma^z$ term responsible for phase noise), but in our case the presence of a nontrivial, continuously applied Hamiltonian $H \of{t}$ means that to change the quantum state, the phase noise must induce transitions between states of different energy under $H \of{t}$, at a finite energy cost\,\cite{Martinis2003}. This finite energy cost eliminates the low frequency divergence in the noise power spectrum $S \of{\omega}$ and dramatically suppresses phase noise, leading to our circulation pattern best fit by assuming a white noise $T_{2} \simeq 2 T_{1}$, the standard limit from photon losses. We note that this replicates the result of Averin \textit{et al.}\,\cite{Averin}, who demonstrated phase noise suppression by applying sequential SWAP operations in a ring of qubits; our continuous Hamiltonian can be thought of as a passive, analog equivalent to their gate-based method when viewed through the lens of quantum error suppression.

\begin{figure*}
\begin{centering}
\includegraphics[width=160mm]{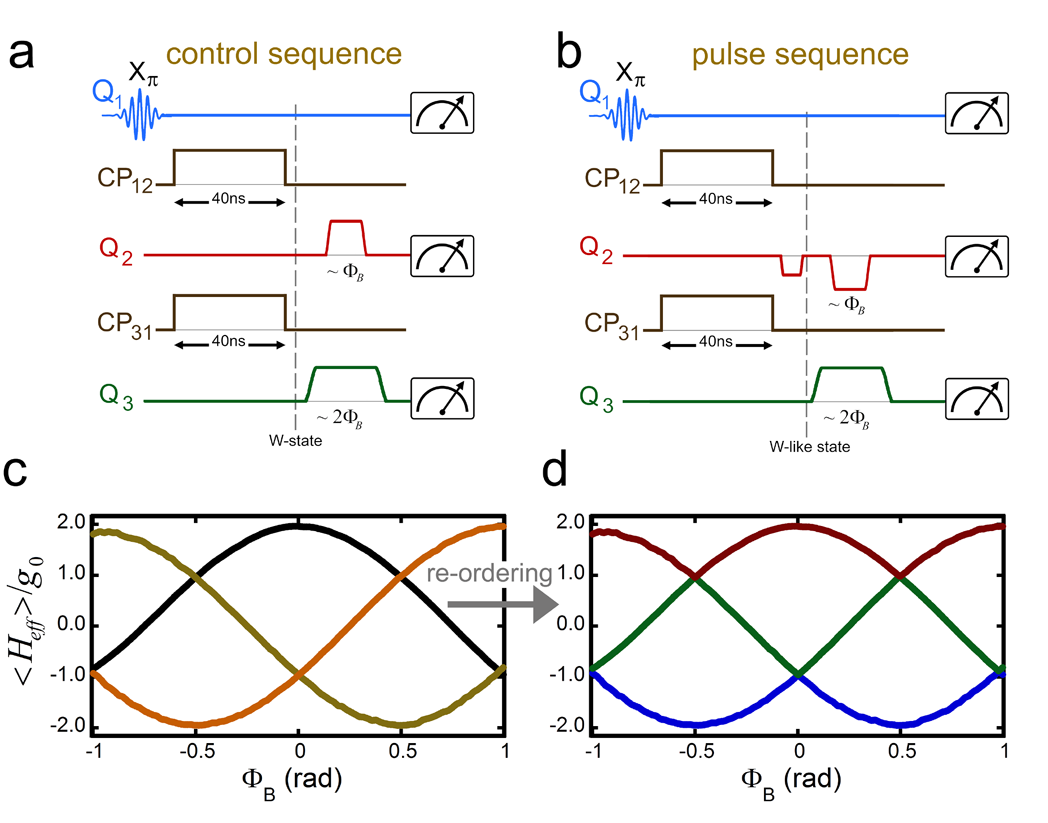}
\par\end{centering}
\caption{ \textbf{Generating the full spectrum of eigen-energies}. The pulse sequence used for generating all energy eigenstates of the system. \textbf{(a)} By exciting one qubit and setting its coupling to other qubits to the fixed value of $g_0/2\pi=4$\,MHz for $40$ns, a $W$-state of the three qubits is generated. After that, individual qubits are rotated to produce the desired phases for creating the eigen-states of the problem according to the lemma. \textbf{(b)} $W$-like states are created, by a protocol similar to \textbf{(a)}. Next, proper phases were given to each qubit to create all the states on the energy manifold. A similar protocol (not shown) to \textbf{(a)} and \textbf{(b)} was used to create eigen-states on the third manifold. \textbf{(c)} After creating the eigen-energies of the system, the full density matrix of the system was measured, and the expectation values of energy for all eigen-energies extracted. \textbf{(d)} The three protocols are insensitive to degeneracies and only provide the eigen-states that are connected to each other through infinitesimal change of $\Phi_B$. From panel \textbf{(c)} the degeneracies of the Hamiltonian become visible; here, we rearrange the measured values shown in \textbf{(c)} and re-plot them. }
\label{fig:fit}
\end{figure*}

\vspace{20mm}

\vspace{5mm}

\psection{Chirality.} The energy spectra provide a holistic picture that allows exploring quantum correlations in various eigenstates of the system. In particular, measuring the chirality of can provide insight into how states on different energy manifolds respond to gauge fields. The chirality operators is defined as

\begin{equation}\label{eq:Chi}
\hat{\chi}=\overrightarrow{\sigma}_{Q_1}.(\overrightarrow{\sigma}_{Q_2}\times\overrightarrow{\sigma}_{Q_3}),
\end{equation}
where $\overrightarrow{\sigma}=(\sigma^X,\sigma^Y,\sigma^Z)$.  Chirality is computed using the measured density matrices ($tr(\hat{\rho}\hat{\chi})$), and is presented in Fig. S5\textbf{(b)}. On the ground and first-excited manifolds, any non-zero value of $\Phi_B$ breaks TRS and leads to chiral states, with the chirality. Close to $\Phi_B=0$, The highest excited manifold, shows a weaker dependance on $\Phi_B$.

\begin{figure}
\begin{centering}
\includegraphics[width=155mm]{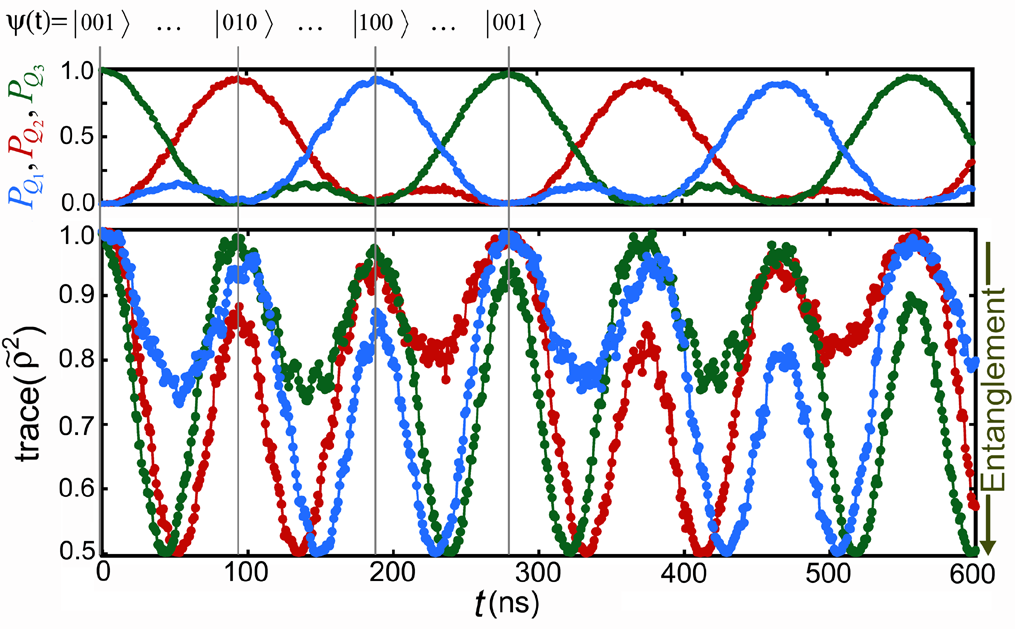}
\par\end{centering} \caption{ \textbf{Entanglement Dynamics during circulation}. The top panel shows the photon occupation probability of each qubit $P_{Q_j}$ as a function of time. Data is similar to Fig.\,2\textbf{(c)} of the main text and presented here for the ease of comparison with the entanglement measurements. At time $t=0$, the system is prepared in the $|001\rangle$ state, which has zero entanglement between the qubits. At later times $\langle\sigma^{X}\rangle$, $\langle\sigma^{Y}\rangle$, and $\langle\sigma^{Z}\rangle$ of each qubit are measured. From the expectation values of these Pauli operators, the reduced density matrix of each qubit $\tilde{\rho}$ was constructed and the entanglement of each qubit with the others computed and presented in the lower panel.}
\label{fig:entanglement}
\end{figure}

\vspace{8mm}
\psection{A single photon circulator.} From the quantum technology perspective, the setup presented here is a single photon circulator device and is interesting by itself. However, the excitation circulation in this device is distinct from circulations seen in classical non-reciprocal three port devices\,\cite{Fang2012,Fang2013,Estep2014,Lehnert2015}. To gain deeper insight into the underlying circulation mechanism, we investigate the role of quantum correlations during the photon circulation. The inherent quantum nature of the circulation observed here manifests itself in the generation of entanglement among the qubits during the evolution. In Fig.\,S6, we measure the reduced density matrix $\tilde{\rho}$ of each qubit for the single-photon circulation protocol. When a qubit is not entangled with other qubits, its $tr(\tilde{\rho}^{2})$ is maximized to one, and when it is fully entangled with other qubits, its $tr(\tilde{\rho}^{2})$ is minimized to $0.5$. Comparing the top and lower panels provides insight into how the excited state circulates among the qubits. If an excitation moves from $Q_{j}$ to $Q_{k}$, then these two qubits become entangled. This can be seen from comparing minima in the lower panel with maxima in the top panel and their successional appearances. When the excitation reaches the second qubit, all qubits become disentangled (gray vertical lines). During the passage of excitation between two qubits, the third qubit also becomes partially entangled with them. Therefore, as the excitation circulates, entanglement among the three qubits is periodically generated and annihilated. A time-resolved measurement of the the full density matrix of the system during the circulation is also presented.

\begin{figure*}
\begin{centering}
\includegraphics[width=165mm]{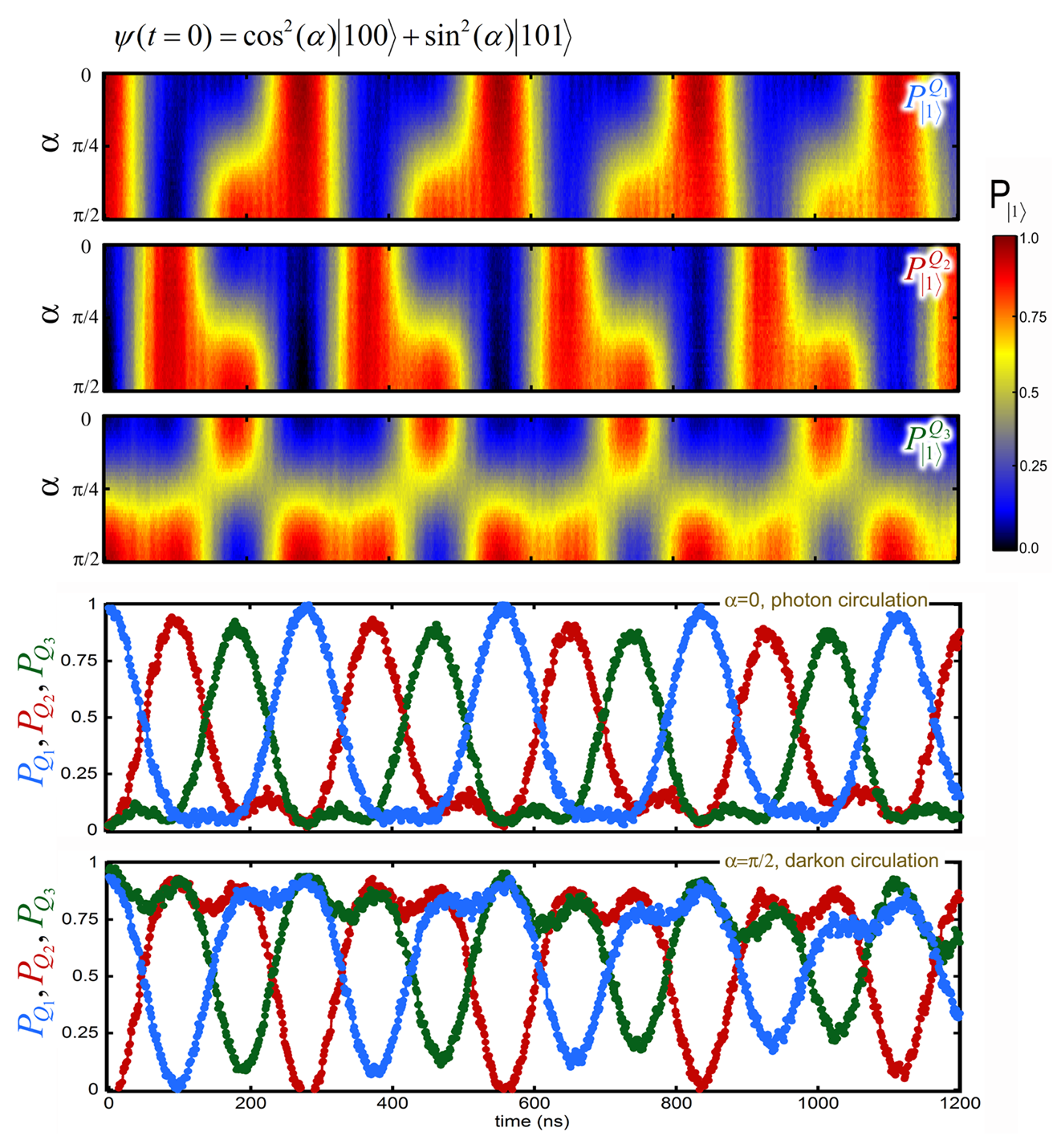}
\par\end{centering}
\caption{ \textbf{Excitation circulation in the superposition of single- and two-photon manifolds.} The top three panels show the measured $P_{Q}$ for the three qubits as a function of time and initial state. The initial state was gradually varied from the single photon manifold when $\alpha=0$ to be in the two-photon manifold when $\alpha=\pi/2$. As discussed in the main text, the counter propagating nature of the two-photon circulation makes it more convenient to consider the absence of photon circulation, hence the name "darkon". The lower panels show horizontal cuts to the data at $\alpha=0$ and $\alpha=\pi/2$, which are similar to what is shown in Fig.(2) of the main text. Considering the qubit's layout, for single excitation ($\alpha=0$), the excitation circulates in a clockwise direction; for two excitations ($\alpha=\pi/2$), one can see a counter-clockwise  circulation. For other values of $\alpha$ a superposition is created. At $\alpha=\pi/4$, the excitation does not circulate and goes back and forth between $Q_1$  and $Q_2$. This can be seen in the dominance of the yellow color when $\alpha=\pi/4$ in the $Q_3$ panel.}
\label{fig:fit}
\end{figure*}

\begin{figure*}
\begin{centering}
\includegraphics[width=130mm]{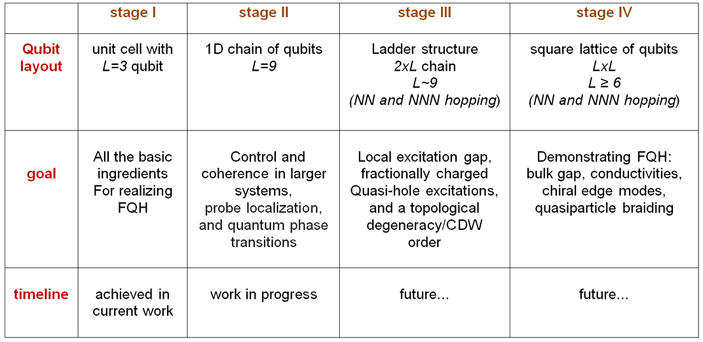}
\par\end{centering}
\caption{ \textbf{A road-map toward realization of FQH states with superconducting qbutis.} This table shows the road-map that we are considering for synthesizing FQH states with superconducting circuits and our progress in this path.}
\label{fig:roadmap}
\end{figure*}

\begin{figure*}
\begin{centering}
\includegraphics[width=145mm]{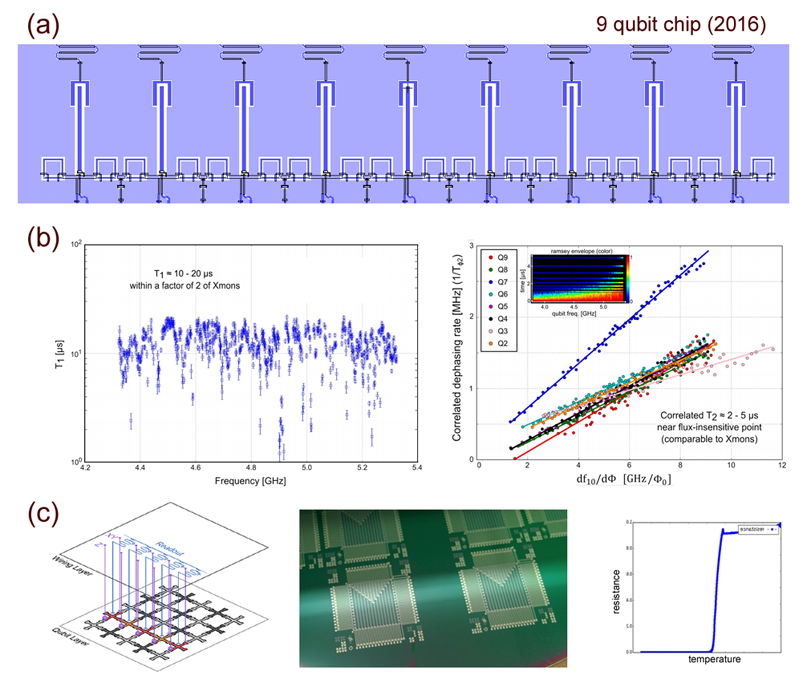}
\par\end{centering}
\caption{ \textbf{Experimental progress toward making FQH states with qubits}. Larger lattices with good coherent times are needed for synthesizing FQH states. We break down the FQH requirements to three experimental challenges: (1) controlling larger superconducting qubits lattices, (2) reducing the energy decay and decoherence in large lattices, and (3) laying out 2D architecture. \textbf{(a)} On the stage II of this project, we scaled up and placed 9 g-mon quits in a 1D chain. At the core of the experimental challenge, in going from 3 to 9 qubits, was to make the layout such the full control over the qubits and their coupling is maintained and the cross-talk between various elements is kept to minimum. The 9-qubit chip was fabricated and after a single iteration, the desired control is achieved, and the cross-talk has been maintained at the tolerable level of less than one percent. \textbf{(b)} Energy decay (left) and decoherence (right) studies of the 9 qubit chip. A key effort in our lab is to improve the coherence of the chips. While it is easy to have very good coherence properties at the level of a few qubits or when the qubits are not connected, it is rather challenging to maintain coherence performance as the system size gets larger and the qubits are coupled with adjustable couplers. Through careful design and utilizing the latest advancements in our micro-fabrication techniques, we were able to show comparable, or even sometimes better, performance in the coherence measurement of the 9-qubit chip compared to the 3 qubit chip sample. These preliminary results clearly shows the feasibility of scaling up with superconducting devices to the desired level for realizing FQH physics. \textbf{(c)} To realize FQH states we need either at least a $6$ by $6$ lattice. Therefore, the capability to lay out 2D lattices is required. Here, we show our scheme and preliminary results in how to create 2D lattices. The idea (left) is that one place the qubits on chip and the control lines and readout resonators on a separate chip, and flips the second chip over the first chip. The schematic is adopted from reference\,\cite{RamiNature2014}. Some of the main challenge in this approach is to assure uniformity in the separation of the two chips and electrical connectivity of the two chips, forming one circuit with a common ground. The middle panel shows the test chip that we fabricated for this purpose. The signature of proper electrical connectivity of the two layers, would be the appearance of superconductivity and dropping of the resistance to zero in a current. The current path starts from one chip, goes to the other chip, and comes back to the first chip. As can be seen in the right panel, we were able to successfully demonstrate superconductivity recently.}
\end{figure*}